\documentclass[twocolumn,showpacs,amsmath,amssymb,prl]{revtex4}% Physical Review B
\usepackage{epsf}
\usepackage{dcolumn}% Align table columns on decimal point
\usepackage{bm}% bold math
\everymath{\displaystyle}
\begin{document}

\title{Scalar particle in general inertial and gravitational
fields and conformal invariance revisited}
\author{Alexander J. Silenko}
\affiliation{Research Institute for Nuclear Problems, Belarusian State University, Minsk 220030, Belarus\\
Bogoliubov Laboratory of Theoretical Physics, Joint Institute for Nuclear Research,
Dubna 141980, Russia}

\date{\today}

\begin {abstract}
The new manifestation of conformal invariance for a
massless scalar particle in a %%%non-Minkowskian
Riemannian spacetime of general relativity is found. Conformal transformations conserve the Hamiltonian and wave function in the Foldy-Wouthuysen representation.
Similarity of manifestations of conformal invariance for massless scalar and Dirac particles is proved.
New exact Foldy-Wouthuysen Hamiltonians are derived for both massive and massless scalar particles in a general
static spacetime %%%%%%%%%%%%metric instead of spacetime
%%%inertial and gravitational fields
and %%%%%%%%%%%%for particles
in a frame rotating in the Kerr field approximated by a spatially isotropic metric.
%%%%%%%%%%%%the field of a rotating source.
The latter case covers an observer on the ground of the Earth or on a satellite and
takes into account %not only the rotation but also
the Lense-Thirring effect.
High-precision formulas are obtained for an arbitrary spacetime metric.
General quantum-mechanical equations of motion are derived. Their classical limit coincides with corresponding
classical equations.
\end{abstract}

%\pacs {03.65.Pm, 04.62.+v, 11.10.Ef, 11.30.-j}
% 03.65.Pm  Relativistic wave equations
% 04.62.+v  Quantum fields in curved spacetime
% 11.10.Ef  Lagrangian and Hamiltonian approach
% 04.20.Jb  Exact solutions
% 11.30.-j  Symmetry and conservation laws
\maketitle

\section{Introduction}

Penrose \cite{Penrose} has discovered fifty years ago the conformal invariance of the
covariant Klein-Gordon (KG) equation \cite{KG} for a
massless scalar particle in a Riemannian spacetime added by
an appropriate term describing a nonminimal coupling to the scalar
curvature. Chernikov and Tagirov
\cite{CheTagi} have given clear explanations of this wonderful
result. Their study involved the case of a nonzero mass and
$n$-dimensional Riemannian spacetime. The inclusion of the
Penrose-Chernikov-Tagirov term has been argued for both massive and massless particles
\cite{CheTagi}.
The next step in investigation of the problem of conformal invariance
of the KG equation has been made by Accioly and Blas
\cite{AccBlas}. They have performed the exact %%%Foldy-Wouthuysen
Foldy-Wouthuysen (FW) transformation for a massive spin-0 particle in static
spacetimes and have found new telling arguments in favor of the predicted %%%PCT
coupling to the scalar curvature. A derivation of the relativistic
FW Hamiltonian %fulfilled in Ref. \cite{AccBlas}
is important for a comparison of gravitational (and inertial) effects for
scalar and Dirac particles. However, the transformation method used in Ref. \cite{AccBlas} is
inapplicable to massless particles. In addition, it cannot be applied for
nonstatic spacetimes. This does not allow us to obtain 
information about a %specific
manifestation of the conformal
invariance in the FW representation.

In the present work, we consider a scalar particle in arbitrary
spacetimes
%%%
in the framework of general relativity (GR). To obtain a Hamiltonian form of the initial covariant
KG equation not only for massive particles but also for massless
ones, we use the generalization of the Feshbach-Villars
transformation \cite{FV} proposed in Ref. \cite{TMP2008}. Then we
fulfill the %subsequent
FW transformation and prove the conformal invariance of the
relativistic FW Hamiltonian for a wide class of
%static and stationary
inertial and gravitational fields. We derive general
quantum-mechanical equations of motion and obtain their classical limit.
%As an example, quantum-mechanical description of the Lense-Thirring effect is
%presented and its classical limit is found.

We denote world and spatial indices by greek and latin letters
$\alpha,\mu,\nu,\ldots=0,1,2,3,~i,j,k,\ldots=1,2,3$, respectively. Tetrad
indices are denoted by latin letters from the beginning of the
alphabet, $a,b,c,\ldots = 0,1,2,3$. %%%%%%%%%%%%Separate
Temporal and spatial tetrad indices are
distinguished by hats. The signature is $(+---)$, the Ricci scalar
curvature is defined by
$R=g^{\mu\nu}R_{\mu\nu}=g^{\mu\nu}R^\alpha_{~\mu\alpha\nu}$, where
$R^\alpha_{~\mu\beta\nu}=\partial_\beta\Gamma^\alpha_{~\mu\nu}-\ldots$
is the Riemann curvature tensor. We use the system of units
$\hbar=1,~c=1$ except for some specific expressions.

\section{Importance of the Penrose-Chernikov-Tagirov term} %for massive scalar particles}

The covariant KG \cite{KG}
equation with the additional term \cite{Penrose,CheTagi} describes a scalar particle in
a Riemannian spacetime and is given by
\begin{equation}
(\square+m^2-\lambda R)\psi=0,~~~
\square\equiv\frac{1}{\sqrt{-g}}\partial_\mu\sqrt{-g}g^{\mu\nu}\partial_\nu.
\label{eqKG} \end{equation} Minimal (zero) coupling corresponds
to $\lambda=0$, while the Penrose-Chernikov-Tagirov coupling is defined by $\lambda=1/6$
\cite{footnote}. For noninertial (accelerated and rotating)
frames, the spacetime is flat and $R=0$.

For massless particles, the conformal transformation
\begin{equation}
\widetilde{g}_{\mu\nu}=O^{-2}g_{\mu\nu}\label{conftrf}
\end{equation} conserves %remains
%\widetilde{g}_{\mu\nu}=\widetilde{\Omega}^{-2}g_{\mu\nu}
%unchanged
the form of Eq. (\ref{eqKG}) but changes the operators and the wave
function \cite{Penrose,CheTagi}:
\begin{equation}
(\widetilde{\square}-\frac16 \widetilde{R})\widetilde{\psi}=0,~~~
\widetilde{\psi}=O\psi. \label{eqcin}
\end{equation} In Ref. \cite{CheTagi}, higher dimensionality was
also considered.

The corresponding classical equation
$$ %\begin{equation}
g^{\mu\nu}p_\mu p_\nu-m^2=0 $$ %\label{eqcla}\end{equation}
is also conformal for a massless particle. It does
not contain any nonminimal coupling to the scalar curvature.
Therefore, the square of the classical momentum corresponds to the
operator $-\hbar^2(\square-R/6)$ \cite{CheTagi}.

Chernikov and Tagirov \cite{CheTagi} have shown the importance of the
additional term for massive particles. They have proved that the
requirement for motion to be quasiclassical for a large momentum
is satisfied for massive and massless particles only when
$\lambda=1/6$. This choice of $\lambda$ has been additionally
%%%grounded
substantiated in Refs. \cite{Faraoni,GribPbr}.

An important development of problem of the Penrose-Chernikov-Tagirov coupling in the GR for
\emph{massive} particles has been made by Accioly and Blas
\cite{AccBlas}. They analyzed a dependence of the form of the
FW Hamiltonian on the value of $\lambda$ and considered the diagonal static metric
%in isotropic coordinates:
\begin{equation}
ds^2={V(\bm r)}^2 (dx^0)^2-{W(\bm r)}^2 (d\bm r)^2 \label{tamet}
\end{equation} with arbitrary $V(\bm r),W(\bm r)$. The choice of the metric
allowed an \emph{exact} FW transformation by the method used in Ref. \cite{AccBlas}.
This method included the Feshbach-Villars transformation (inappropriate for
massless particles) in order to bring
the initial equation (\ref{eqKG}) to the Hamiltonian form.
Next, nonunitary and FW transformations resulted in the FW
Hamiltonian \cite{AccBlas}:
\begin{equation}
{\cal H}_{FW}\!=\!\rho_3\sqrt{m^2V^2\!+\!{\cal F}\bm p^2{\cal
F}\!-\!\frac14\nabla{\cal F}\!\cdot\!\nabla{\cal F}\!+\!{\cal
D}_\lambda(V,W)}, \label{FWHam}
\end{equation}
where $\bm p=-i\nabla$ is the momentum operator and
$\rho_i~(i=1,2,3)$ are the Pauli matrices. Only for $\lambda=1/6$, the Darwin term ${\cal
D}_\lambda(V,W)$ has the simple form and is equal to ${\cal
F}\Delta{\cal F}/6$ \cite{AccBlas}.

However, the important result obtained by Accioly and Blas \cite{AccBlas}
demonstrates only a shadow of the conformal invariance, because it
does not cover the case of $m=0$. %Unlike Ref. \cite{AccBlas}, we
We perform general examination of the problem.

\section{Generalized Feshbach-Villars transformation}

The general form of the covariant KG equation reads
\begin{equation} \begin{array}{c} \left(\partial_0^2+\frac{1}{g^{00}\sqrt{-g}}
\left\{\partial_i,\sqrt{-g}g^{0i}\right\}\partial_0 \right.\\
\left.+\frac{1}{g^{00}\sqrt{-g}}\partial_i\sqrt{-g}g^{ij}\partial_j+\frac{m^2-\lambda
R}{g^{00}}\right)\psi=0.
\end{array} \label{eq1i} \end{equation}
The curly bracket $\{\ldots,\ldots\}$ denotes the anticommutator.

There is an ambiguity \cite{Mosta} in the definition of the parameter of
the Feshbach-Villars transformation. We use the generalized
Feshbach-Villars transformation proposed in Ref.
\cite{TMP2008} and based on this ambiguity. In the considered
case, the transformation consists in the following definition of components of the
wave function:
\begin{equation} \begin{array}{c} \psi=\phi+\chi, ~~~
i\left(\partial_0+\Upsilon\right)\psi= N(\phi-\chi),\\
%\label{eq3i}\end{equation}
%\begin{equation}
\Upsilon= \frac{1}{2g^{00}\sqrt{-g}}\left\{\partial_i,\sqrt{-g}g^{0i}\right\},
%\label{eq4i}\end{equation}
\end{array} \label{eq3i} \end{equation} where $N$ is an arbitrary nonzero real
parameter. For the Feshbach-Villars transformation, it is definite
and equal to the particle mass $m$. This generalization allows us to represent
Eq. (\ref{eq1i}) in the Hamiltonian form describing both massive and
massless particles:
\begin{equation} \begin{array}{c}  i\frac{\partial\Psi}{\partial t}\!=\!{\cal H}\Psi,
~~~ {\cal
H}\!=\!\rho_3
\frac{N^2\!+\!T}{2N}\!+\!i\rho_2 \frac{-N^2\!+\!T}{2N}\!-\!i\Upsilon,\\
 T=\frac{1}{g^{00}\sqrt{-g}}\partial_i\sqrt{-g}g^{ij}\partial_j+\frac{m^2-\lambda R}{g^{00}}
-\Upsilon^2.
\end{array} \label{eq5i} \end{equation}
%Equation (\ref{eq5i}) describes both massive and massless particles.

Similarly to Ref. \cite{AccBlas}, then we perform the nonunitary
transformation $\Psi'=f\Psi$ to obtain a pseudo-Hermitian (more exactly,
$\rho_3$-pseudo-Hermitian) Hamiltonian:
%$$\Psi'=f\Psi, ~~~
${\cal
H}'=f{\cal H}f^{-1}, ~ {\cal H}'=\rho_3{{\cal H}'}^\dag\rho_3.$
In the case under consideration,
\begin{equation} \begin{array}{c} f=\sqrt{g^{00}\sqrt{-g}},~~~ \Upsilon'=\frac{1}{2f}\left\{\partial_i,\sqrt{-g}g^{0i}\right\}\frac{1}{f}, \\
%T'=\frac{1}{\sqrt{g^{00}\sqrt{-g}}}\partial_i\sqrt{-g}g^{ij}\partial_j\frac{1}{\sqrt{g^{00}\sqrt{-g}}}
%+\frac{m^2-\lambda R}{g^{00}} -(\Upsilon^2)', \\
%(\Upsilon^2)'=(\Upsilon')^2=-\frac14\left(
%\frac{1}{\sqrt{g^{00}\sqrt{-g}}}\left\{\partial_i,\sqrt{-g}g^{0i}\right\}\frac{1}{\sqrt{g^{00}\sqrt{-g}}}\right).
T'=\frac1f\partial_i\sqrt{-g}g^{ij}\partial_j\frac1f
+\frac{m^2-\lambda R}{g^{00}} -(\Upsilon^2)'.
\end{array}\label{eqfrf}\end{equation} Transformed operators are
denoted by primes and $(\Upsilon^2)'\!=\!(\Upsilon')^2$.
Tedious but simple calculations result in
\begin{equation} \begin{array}{c}  {\cal H}'=\rho_3 \frac{N^2+T'}{2N}+i\rho_2
\frac{-N^2+T'}{2N}-i\Upsilon',\\
%\end{array} \label{eq7i} \end{equation}
%\begin{equation} \begin{array}{c}
T'=\partial_i\frac{G^{ij}}{g^{00}}\partial_j +\frac{m^2-\lambda
R}{g^{00}}+
\frac1f\nabla_i\left(\sqrt{-g}G^{ij}\right)\nabla_j\left(\frac1f\right)\\
+\sqrt{\frac{\sqrt{-g}}{g^{00}}}G^{ij}\nabla_i\nabla_j\left(\frac1f\right)
+\frac{1}{4f^4}\left[\nabla_i(\Gamma^i)\right]^2\\-\frac{1}{2f^2}
\nabla_i\left(\frac{g^{0i}}{g^{00}}\right)\nabla_j\left(\Gamma^j\right)-\frac{g^{0i}}{2g^{00}f^2}
\nabla_i\nabla_j\left(\Gamma^j\right), \\
\Upsilon'\!=\!\frac12\!\left\{\partial_i,\frac{g^{0i}}{g^{00}}\right\},
~~~ G^{ij}\!=\!g^{ij}\!-\!\frac{g^{0i}g^{0j}}{g^{00}},~~~\Gamma^i\!=\!\sqrt{-g}g^{0i},
\end{array}\label{eqfr2}\end{equation}
where the nabla operators act only on the operators in brackets.
%and the primes denote transformed operators.
Equation
(\ref{eqfr2}) is exact and covers any inertial and
gravitational fields.

\section{Foldy-Wouthyusen transformation}

General methods of the FW transformation for relativistic
particles have been developed in Refs. \cite{JMP,PRA}. They belong
to step-by-step methods performing the transformation as a result
of subsequent iterations. We use the version \cite{TMP2008}
adapted to scalar particles.
%
%The GFV Hamiltonian can be decomposed into operators commuting
%(${\cal M}$ and ${\cal E}$) and anticommuting (${\cal O}$) with
%$\rho_3$: 5\begin{equation} \begin{array}{c} {\cal H}'=\rho_3
%{\cal M}+{\cal E}+{\cal O}, ~~~\rho_3 {\cal M}={\cal M}\rho_3,\\
%\rho_3 {\cal E}={\cal
%E}\rho_3, ~~~\rho_3 {\cal O}=-{\cal O}\rho_3,\\
%{\cal M}=\frac{N^2+T'}{2N}, ~~~ {\cal E}=-i\Upsilon', ~~~ {\cal
%O}=i\rho_2 \frac{-N^2+T'}{2N}.
%\end{array}
%\label{eqH} \end{equation}
%
In this case, the relativistic FW transformation is
carried out with the $\rho_3$-pseudounitary operator
($U^\dag=\rho_3U^{-1}\rho_3$) \cite{TMP2008}
%\begin{equation}
%U=\frac{\epsilon+{\cal M}+\rho_3{\cal
%O}}{\sqrt{2\epsilon(\epsilon+{\cal M})}}, ~~~\epsilon=\sqrt{{\cal
%M}^2+{\cal O}^2}. \label{eq9i} \end{equation} In the considered
%case,
\begin{equation}
U=\frac{\epsilon+N+\rho_1(\epsilon-N)}{2\sqrt{\epsilon N}},
~~~\epsilon=\sqrt{T^\prime}. \label{eq8i}
\end{equation}

It is important that the Hamiltonian obtained as a result of the
transformation does not depend on $N$ \cite{TMP2008}. %:
%\begin{equation} \begin{array}{c}
%{\cal H}^{(1)}\!=\!\rho_3\epsilon\!+\!{\cal E}^{(1)}\!+\!{\cal
%O}^{(1)},~~~[\rho_3,{\cal E}^{(1)}]\!=\!\{\rho_3,{\cal O}^{(1)}\}\!=\!0,\\ {\cal
%E}^{(1)}\!=\!-i\Upsilon'\!+\!\frac{1}{2\sqrt{\epsilon}}\left[{\sqrt{\epsilon},
%[\sqrt{\epsilon}},{\cal G}]\right]\frac{1}{\sqrt{\epsilon}}, \\
%{\cal O}^{(1)}\!=\!\rho_1\frac{1}{2\sqrt{\epsilon}}[\epsilon,{\cal
%G}]\frac{1}{\sqrt{\epsilon}},~~~ {\cal G}=-i\partial_0-i\Upsilon'.
%\end{array}\label{eq12i}\end{equation}
This shows a self-consistency of the used transformation method.
Next transformation \cite{TMP2008} eliminates residual odd terms and leads to the final form of the
\emph{approximate} relativistic FW Hamiltonian:
\begin{equation} \begin{array}{c}
{\cal H}_{FW}\!=\!\rho_3\epsilon\!-\!i\Upsilon'\!-\!\frac{1}{2\sqrt{\epsilon}}\!\left[\sqrt{\epsilon},
\left[\sqrt{\epsilon},(i\partial_0\!+\!i\Upsilon')\right]\right]\!\frac{1}{\sqrt{\epsilon}}.
\end{array} \label{eqf} \end{equation}
%It can be shown that the last term in
%Eq. (\ref{eqf}) is of the second or higher orders in the Planck
%constant and of the second order in field parameters. All
%relativistic step-by-step methods are approximate
%\cite{EK,JMPcond} and cannot usually provide so high accuracy.
%Nevertheless, they perfectly describes the contact (Darwin)
%interaction and strong field effects (see Refs.
%\cite{PRD,OSTRONG}). Allowance for the last term in Eq.
%(\ref{eqf}) exceeds the achievable accuracy. Further, this term will be omitted.

\section{Exact Foldy-Wouthuysen transformation and conformal invariance}
%new manifestation of conformal invariance}

The used method ensures the exact FW transformation for
%any static spacetimes and
a wide class of spacetime metrics.
%stationary ones including a general noninertial frame.
%Since the method is suitable for massless particles, the
The manifestation of conformal invariance can also be investigated in
detail.

The sufficient condition of the exact FW transformation
\cite{JMP,PRA,TMP2008} applied to scalar particles is given by $\partial_0T^\prime-[T^\prime,\Upsilon']=0$.
%\begin{equation}
%[T^\prime,{\cal G}]=i
%%%$$
%%%\partial_0T^\prime-[T^\prime,\Upsilon']=0. $$
%\label{eFW} \end{equation}
When it is satisfied, the exact FW Hamiltonian reads
\begin{equation} {\cal H}_{FW}=\rho_3\sqrt{T^\prime}- i\Upsilon'.
\label{eqFW} \end{equation} Equation (\ref{eqFW}) covers \emph{all
static spacetimes} ($\Upsilon'=0$) and some important cases of stationary ones.
%%%For static spacetimes, $\Upsilon'=0$.

Since general expressions for the scalar Ricci curvature are very
cumbersome, we restrict ourselves to an analysis of several
special cases. % First,
For the metric defined by Eq.
(\ref{tamet}), the result of our calculations formally coincides with
Eq. (\ref{FWHam}). However, the case of $m=0$ can now be considered. The explicit expression for ${\cal
D}_\lambda(V,W)$ \cite{AccBlas} shows the presence of conformal invariance for massless particles if and only if $\lambda=1/6$. In this case, conformal transformation
(\ref{conftrf}) does not change the FW Hamiltonian and the FW
wave function $\Psi_{FW}$. These
manifestations of conformal invariance radically differ from
those for the covariant KG equation and the corresponding
wave function. %$\psi$.

The validity of the found properties can be checked for the scalar
particle in nonstatic spacetimes.
%In particular, the exact FW
%transformation can be fulfilled for
The metric of the rotating Kerr source has been reduced to the
Arnowitt-Deser-Misner form \cite{ADM} by Hergt and Sch\"afer
\cite{hergt}. This form reproduces the Kerr
solution only approximately. %The FW transformation
%for the scalar particle in the field of the rotating source can be
%significantly simplified in the approximation admitting
The form of the metric can be additionally simplified due to %after
an introduction of spatially isotropic coordinates and
dropping terms violating the isotropy \cite{OSTRONG}:
\begin{equation} \begin{array}{c}
ds^2 = V^2(dx^0)^2 - W^2\delta_{ij}(dx^i - K^idx^0)(dx^j\\ -
K^jdx^0),  ~~~ \bm K=\bm\omega\times\bm r. \end{array}
\label{Ltisotr} \end{equation} The use of the approximate Kerr metric allows us to fulfill the
\emph{exact} FW transformation when $V,W$, and $\bm{\omega}$ depend only on the isotropic radial
coordinate $r$. In this approximation, the metric %of the rotating Kerr source takes the form
is defined by
\begin{equation}
\begin{array}{c}   V(r)\!=\!\frac{\kappa_-}{\kappa_+}
\! + \!{\cal O}\!\left(\frac{\mu a^2}{r^{3}}\right), ~~~
 W(r)\!=\! \kappa_+^{2}\! +\! {\cal O}\!\left(\frac{\mu a^2}{r^{3}}\right), \\
\bm{\omega}(r)= {\frac{2\mu c}{r^3}}\,\bm{a}\left[1 - {\frac
{3\mu}{r}} + {\frac {21\mu^2}{4r^2}}+{\cal
O}\!\left(\frac{a^2}{r^{2}}\right)\right].
\end{array} \label{Okerr}
\end{equation}
Here $\kappa_\pm=1\pm\mu/(2r),~\bm{a} = \bm{J}/(Mc),~\mu = GM/c^2$; the total mass $M$ and
the total angular momentum $\bm{J}$ (directed along the $z$ axis)
define the Kerr source uniquely. The leading term in the
expression for $\bm{\omega}(r)=\omega(r)\bm e_z$ corresponds to
the Lense-Thirring approximation.

We can pass on from the Kerr field approximated by Eqs. (\ref{Ltisotr}) and (\ref{Okerr}) to a %noninertial
frame rotating in this field with the angular velocity $\bm o$
after the transformation $dx^i\rightarrow d{X}^i=dx^i\!+\!(\bm
o\times\bm r)dx^0$ \cite{footn}. The stationary metric of this frame can be
obtained from Eqs. (\ref{Ltisotr}) and (\ref{Okerr}) with the replacement
$\bm\omega\rightarrow \bm\Omega\!=\!\bm\omega\!-\!\bm o$. In particular, it covers an
observer on the ground of a rotating source like the Earth or
on a satellite. In this case, $\bm
o=\bm{J}/I$, where $I$ is the moment of inertia. The exact FW Hamiltonian is given by Eq.
(\ref{eqFW}) where
\begin{equation}\begin{array}{c} T'=m^2V^2+{\cal F}\bm p^2{\cal
F}-\frac14\nabla{\cal F}\cdot\nabla{\cal F}+{\cal
D}_\lambda(V,W)\\+\frac\lambda2(x^2+y^2)(\Omega'_r)^2, ~~~ {\cal
D}_\lambda(V,W)=\lambda{\cal F}\Delta{\cal
F}\\+\left(1\!-\!6\lambda\right)\frac{V}{2W^2}\left[{\cal
F}\left(\frac{2W'_r}{r}\!+\!W''_{rr}\right)\!+\!\frac{2V'_r}{r}\!+\!V''_{rr}\right],\\
%%%- i\Upsilon'=-\bm\Omega\cdot(\bm r\times\bm p),
- i\Upsilon'=\bm\Omega\cdot(\bm r\times\bm p), \label{eqme}
\end{array} \end{equation}
%Here $\bm L$ is the operator of angular momentum
and derivatives with respect to $r$ are denoted by indices.
%For the rotating frame $\Omega(r)=const,~V(r)=W(r)=1$;
In particular, for the Lense-Thirring metric
%%%$\Omega(r)=2GJ/r^3,~V(r)=1-GM/r,~W(r)=1+GM/r$.
$\bm\Omega(r)=2G\bm{J}/r^3,~V(r)=1-GM/r,~W(r)=1+GM/r$.

While the metric, (\ref{Ltisotr}) and (\ref{Okerr}), reproduces the Kerr solution only approximately,
the derivation of the exact FW Hamiltonian corresponding to this metric allows an independent
unambiguous determination of the value of $\lambda$.
If and only if $\lambda=1/6$, then the conformal transformation (\ref{conftrf}) changes
neither $T'$ nor ${\cal H}_{FW},~\Psi_{FW}$. This property is the
same as for the static metric. %Thus, this new
%manifestation of the conformal invariance for massless particles
%takes place for the rather general stationary metric.

\section{General equations of motion}

The equations for the FW Hamiltonian allow us to derive general
quantum-mechanical equations of motion and then obtain their
classical limit ($\hbar\rightarrow0$). The quantum-mechanical
equations of motion defining the force, velocity, and acceleration
read $(p_0\equiv {\cal H}_{FW})$
%are defined by the commutators of the FW Hamiltonian with appropriate operators:
\begin{equation}\begin{array}{c}
F^i\!\equiv\!\frac {dp^i}{dt}\!=\!\frac12\frac {\partial}{\partial
t}\left\{g^{i\mu},p_\mu\right\}\!+\!\frac{i}{2\hbar}\left[{\cal
H}_{FW},\left\{g^{i\mu},p_\mu\right\}\right],\\
{\cal V}^i\equiv\frac {dx^i}{dt}=\frac{i}{\hbar}\left[{\cal
H}_{FW},x^i\right],~~~ {\cal W}^i=\frac {\partial {\cal
V}^i}{\partial t}\!+\!\frac{i}{\hbar}\left[{\cal H}_{FW},{\cal
V}^i\right].
\end{array}\label{velfactorf}\end{equation}
Any commutation adds the factor $\hbar$ as compared with the
product of operators.

%In the nonrelativistic case, the Wentzel-Kramers-Brillouin (WKB)
%approximation can be used when a de Broglie wavelength is smaller
%than a characteristic size of inhomogeneity region of an external
%field. It has been proved in Ref. \cite{JINRLet1} that satisfying
%this condition allows to use the WKB approximation in the
It has been proved in Ref. \cite{JINRLet1} that satisfying
the condition of the Wentzel-Kramers-Brillouin approximation allows us
to use this approximation in the
relativistic case and to obtain a classical limit of the
relativistic quantum mechanics. Determination of the classical
limit reduces to the replacement of operators in the FW
Hamiltonian and quantum-mechanical equations of motion in the FW
representation by the respective classical quantities. The
classical limit of the general FW Hamiltonian is given by
\begin{equation}
H = \left(\frac{m^2 - G^{ij}p_ip_j} {g^{00}}\right)^{1/2} -
\frac{g^{0i}p_i}{{g}^{00}}.\label{clCog}
\end{equation}
It coincides with the classical Hamiltonian derived in Ref.
\cite{Cogn}.
The classical limit of Eq. (\ref{velfactorf}) reads
\begin{equation}\begin{array}{c}
{\cal V}^i=\frac{G^{ij}p_j} {\sqrt{g^{00}(m^2 - G^{ij}p_ip_j)}}+
\frac{g^{0i}}{g^{00}},\\
F^i= %%%p_\mu\dot{ g}^{i\mu}+g^{0i}\frac {\partial H}{\partial t}+
p_\mu\frac {\partial g^{i\mu}}{\partial t}+g^{0i}\frac {\partial H}{\partial t}+
g^{ij}\partial_jH+p_\mu {\cal V}^j\partial_j g^{i\mu}.
\end{array}\label{forceclass}
\end{equation}
It %also
coincides with the corresponding classical equations which
follow from Hamiltonian (\ref{clCog}) and the Hamilton equations.
%$$\frac{dH}{dt}=\frac{\partial H}{\partial t},~~~ \frac{dx^{i}}{dt}=-\frac{\partial H}{\partial p_{i}},~~~
%\frac{dp_{i}}{dt}=\frac{\partial H}{\partial x^{i}}.$$
Thus, the quantum-mechanical and classical equations are in the
best compliance.

For example, the exact metric of a general noninertial frame
characterized by the acceleration $\bm a$ and the rotation $\bm o$
of an observer is defined by $V=1+\bm a\cdot\bm r$, $W=1$, $\bm\Omega=-\bm o$ \cite{HN}.
In this case, the classical limit of the %Hamiltonian and
equations of motion is given by ($\bm p\equiv(-p_1,-p_2,-p_3)$)
\begin{equation}\begin{array}{c}
%H=(1+\bm a\cdot\bm r)\sqrt{m^2+\bm p^2}-\bm o\cdot(\bm r\times\bm p),\\
\mbox{\boldmath ${\cal V}$}= (1+\bm a\cdot\bm r)\frac{\bm p}{\sqrt{m^2+\bm p^2}}-\bm o\times\bm r,\\
\mbox{\boldmath ${\cal W}$}=-\bm a(1+\bm a\cdot\bm r)-2\bm o\times\mbox{\boldmath ${\cal V}$}-\bm o\times(\bm o\times\bm r)\\
+\frac{2\bm a\cdot\mbox{\boldmath ${\cal V}$}+\bm a\cdot(\bm o\times\bm r)}{1+\bm a\cdot\bm r}\left(\mbox{\boldmath ${\cal V}$}+\bm o\times\bm r\right).
\end{array}\label{acclass}
\end{equation} Equation (\ref{acclass}) agrees with the %well-known
classical results \cite{Noninertial}.

\section{Conformal invariance for Dirac and classical particles}

It is important to compare the conformal transformations in the GR for massless scalar, Dirac, and classical particles.
Our analysis shows that the general Hermitian Dirac Hamiltonian for a massless particle in an arbitrary metric in the presence of an electromagnetic field \cite{OSTRONG} is not changed by the transformation
(\ref{conftrf}). The FW transformation operator for particles in strong external fields obtained
in Ref. \cite{PRA} is also conformally invariant in the case of $m=0$. As a result, the Dirac and
FW wave functions, $\psi$ and $\psi_{FW}$, and the FW Hamiltonian
remain unchanged. These properties of Dirac particles are the same as for scalar ones. The
Hamiltonian of massless classical particles is conformally invariant even if its spin-dependent
part defined by Eqs. (3.18) and (4.12) in Ref. \cite{OSTRONG} is taken into account.

In the general case, the transformation of the initial covariant Dirac equation to the Hermitian Hamiltonian form is performed by the nonunitary operator $f_D\!=\!(\sqrt{-g}e_{\widehat{0}}^0)^{1/2}$ \cite{OSTRONG}. Since the transformation (\ref{conftrf}) leads to $\widetilde{f}_D\!=\!O^{-3/2}f_D$, the conformally transformed wave function of the initial covariant Dirac equation, $\widetilde{\Psi}$, reads
\begin{equation}
\widetilde{\Psi}=\widetilde{f_D^{-1}}\widetilde{\psi}=O^{3/2}f_D^{-1}\psi=O^{3/2}\Psi.
\label{Pentr}\end{equation}
While its transformation is similar to that for the scalar particles, the powers of $O$ in Eqs. (\ref{eqcin}) and (\ref{Pentr}) differ.

The second-order wave equation for the Dirac particles in general electromagnetic and gravitational fields derived in Ref. \cite{OSTRONG} includes the term describing a nonminimal
coupling to the scalar curvature $R$. As the definitions of $R$ in Ref. \cite{OSTRONG} and the present work differ in sign, this term corresponds to $\lambda=1/4$.

\section{Conclusions}

The use of the generalized Feshbach-Villars and relativistic FW transformations allows us to describe the both massive and
massless scalar particles in general noninertial frames and gravitational fields.
The present work demonstrates the new manifestation of the conformal
invariance for massless particles. The conformal transformation conserves the FW Hamiltonian and
the FW wave function while it changes the wave function of the initial KG equation. The similar
conclusion is valid for the Dirac particles. The nonminimal coupling to the scalar curvature is
not a unique property of scalar particles.
%The second-order wave equation for the Dirac particles \cite{OSTRONG} also includes the term similar to PCT one but %corresponding to $\lambda=1/4$.

The results obtained in Ref. \cite{AccBlas} and in the present study
allow us to state the general property of conformal symmetry for massive particles.
Conformal transformation (\ref{conftrf}) changes only such terms in the FW Hamiltonian which are proportional to the particle mass $m$.
This property is valid not only for real scalars (Higgs boson) but also for compound ones (zero-spin atoms and nuclei).

Contemporary methods of (pseudo)unitary and non\-uni\-ta\-ry transformations make it possible to
derive new exact FW Hamiltonians for both massive and massless scalar particles (i)
in the general static spacetime
%%%%%%%%%%%%%%%%%%%%%%%%%%%%%%%%%%%%%%%%%%%%% general static spacetimes
and (ii)
%Another exact FW transformation is fulfilled for %the particles
in the frame rotating in the Kerr field approximated by a spatially isotropic metric.
The latter result covers
an observer on the ground of the Earth or on a satellite. It reproduces not only the well-known
effects of the rotating frame but also the Lense-Thirring effect. For an arbitrary metric, high-precision formula
(\ref{eqf}) is obtained.
The classical limit of the derived general quantum-mechanical equations of motion coincides with
corresponding classical equations.

\section*{Acknowledgements}

The author is indebted to E. A. Tagirov for his interest in the present study and valuable %comments and
discussions and Yu. A. Tsalkou for the analytic computer calculation of $R$. The work was supported by the
%BRFFR
Belarusian Republican Foundation for Fundamental Research
(Grant No. $\Phi$12D-002).

%\footnotesize
%\renewcommand{\baselinestretch}{1.4}


\begin{thebibliography}{}
\bibitem{footn} Frames rotating in the isotropic and Cartesian
coordinates are not equivalent.
\end{thebibliography}
\end{document}